\documentclass[10pt, conference, letterpaper]{IEEEtran}
\IEEEoverridecommandlockouts
\usepackage{cite}
\usepackage{amsmath,amssymb,amsfonts}
\usepackage{algorithmic}
\usepackage{graphicx}
\usepackage{subfigure}
\usepackage{textcomp}
\usepackage{xcolor}
\def\BibTeX{{\rm B\kern-.05em{\sc i\kern-.025em b}\kern-.08em
    T\kern-.1667em\lower.7ex\hbox{E}\kern-.125emX}}
\begin{document}

\bstctlcite{IEEEexample:BSTcontrol}

\title{Challenges in Net Neutrality Violation Detection: A Case Study of Wehe Tool and Improvements}

\author{\IEEEauthorblockN{Vinod S. Khandkar and Manjesh K. Hanawal}
	\IEEEauthorblockA{\textit{Industrial Engineering and Operations Research} \\
		\textit{Indian Institute of Technology Bombay, Mumbai, India}\\
		\{vinod.khandkar, mhanawal\}@iitb.ac.in}
}

\maketitle

\begin{abstract}
We consider the problem of detecting deliberate traffic discrimination on the Internet. Given the complex nature of the Internet, detection of deliberate discrimination is not easy to detect, and tools developed so far suffer from various limitations. We study challenges in detecting the violations (focusing on the HTTPS traffic) and discuss possible mitigation approaches. We focus on `Wehe,' the most recent tool developed to detect net-neutrality violations. Wehe hosts traffic from all services of interest in a common server and replays them to mimic the behavior of the traffic from original servers. Despite Wehe's vast utility and possible influences over policy decisions, its mechanisms are not yet validated by others. In this work, we highlight critical weaknesses in Wehe where its replay traffic is not being correctly classified as intended services by the network middleboxes. We validate this observation using a commercial traffic shaper. We propose a new method in which the SNI parameter is set appropriately in the initial TLS handshake to overcome this weakness. Using commercial traffic shapers, we validate that SNI makes the replay traffic gets correctly classified as the intended traffic by the middleboxes. Our new approach thus provides a more realistic method for detecting neutrality violations of HTTPS traffic.

\end{abstract}

\begin{IEEEkeywords}
Net neutrality, traffic differentiation, detection tools
\end{IEEEkeywords}

\section{Introduction}
\label{sec:intro}
Net neutrality is a guiding principle promoting the ``equal" treatment of all packets over the Internet. However, for economic benefits, ISPs may apply traffic differentiation to a specific service, user, content provider, or any other traffic group on the Internet without making any public declaration. It gives rise to a need to have tools that can detect such malicious activities over the Internet. 


Traffic differentiation (TD) detection involves the coalescence of many elements such as end-systems (user-client and server), probing traffic generation, expected network responses, and TD detection algorithms. These are interdependent components or operations. Hence, in developing TD detection tools, one faces challenges such as crafting internet traffic and conditioning measured network response that suits their detection algorithm. Moreover, as HTTPS traffic becomes prevalent, it is unclear what policies the middle-boxes apply for discrimination as payload provides no signatures. Our first goal is to study various challenges in designing a reliable TD detection mechanism for HTTPS traffic.

Several methods have been proposed in the literature to detect Network neutrality violations as documented in recent surveys \cite{nn_survey_2018, nn_survey_2020}. The literature is rich with various forms of discrimination and its detection, like discrimination of content providers \cite{nnvd_cpdiscri_pls}, end-users \cite{nnvd_userdiscri_plr}, specific services like BitTorrent \cite{bttest}. Our interest in this work is discrimination of services, specifically on streaming services (both audio and video) which are potential candidates from description due to their commercial values and high bandwidth requirements. 

Recently several tools are developed to detect discrimination of services like, \textit{NANO} \cite{nano} \textit{ChkDiff} \cite{chkdiff}, \textit{Netpolice} \cite{netpolice}, \textit{Shaperprobe} \cite{shaperprobe},  \textit{Packsen} \cite{packsen}, \textit{Glasnost} \cite{glasnost}, \textit{Bonafide} \cite{bonafide}, and \textit{Wehe} \cite{wehe}. \textit{NANO} is based on passive measurements, and all others are based on either active or differential probes. Wehe is the latest tool that overcomes many of the drawbacks of the other tools. However, due to the Internet's complexity, many issues persist that prevent its use as a reliable tool.

\textit{Wehe} \cite{wehe} follows a client-service architecture in which content of service of interests (Youtube, Netflix, etc.) are copied (or sniffed) from their respective original servers and hosted on a common server (referred to as replay server). To check if a particular service is discriminated, the client accesses the content of that service from the replay server. The server transfers the requested data with timing relations that mimic the original service's data transfer characteristics over the Internet. The client accesses content from the replay server with and without VPN and compares them to ascertain if there was any difference in the quality.

The basic idea of Wehe is that connection over VPN is encapsulated and will not be subjected to any discrimination as the middleboxes cannot classify the traffic correctly. Whereas content accessed without VPN can be classified correctly due to the signatures induced by the timing relations and can be subjected to discrimination policies. Despite the Wehe tool's vast utility and possible influences over policy decisions, its mechanisms are not yet fully validated by other researchers. 

We investigate Wehe's differentiation detection's performance on traffic using HTTPS protocol, which is predominantly used by all services for security reasons and to avoid detection. We noticed that Wehe uses port 80 (HTTP) for all communications. Middleboxes process all the traffic on this port as non-encrypted traffic and may not classify them correctly as they will not see any signatures (recall that replayed traffic is HTTPS). On the other hand, when  Wehe access traffic over port 443 (HTTP),  traffic is not classified as intended traffic as we demonstrate it using a commercial traffic shaper. As middleboxes do not see traffic from the Wehe's replay server as that coming from the original server,  they may not apply intended traffic discrimination policies. Thus Wehe may not detect any deliberate discrimination resulting in false negatives.

To overcome the limitations of Wehe for HTTPS traffic, we propose a new method in which traffic is accessed on port 443, but with a modification in the TLS handshake. In the TLS handshake, we explicitly send the Server Name Indication (SNI) corresponding to the actual service. We demonstrate that the HTTPS traffic get correctly classified by the middleboxes with these modifications. Our method ensures the middle-boxes classify traffic from replay servers in the same way they do from the original servers for each service and apply the same discriminate policies (if any) on traffic from both replay and original servers. Thus, our method makes it possible to detect discrimination of HTTPS traffic in a realistic setting. Our contributions can be summarized as: \begin{enumerate}
	\item We study various challenges associated with detection of discrimination of HTTPS traffic. We present the categorization of these challenges based on their source, e.g., as protocol and operational environment.
	\item We demonstrate limitations of Wehe in detection of discrimination of HTTPS traffic. Specifically, we demonstrate that HTTPS traffic will not be correctly classified in the Wehe setup using a commercial traffic shaper. 
	\item We propose a mechanism in which traffic of services accessed from the replay server is treated as if it originated from the actual servers. Thus traffic from the replay server also gets subjected to the same discriminated policies applied on the actual service. We 
	\item We validate that replay traffic of our mechanism is treated in the same way originating from actual servers using a commercial traffic shaper.
\end{enumerate}

The  paper is organizaed as follows. Sec. \ref{sec:rel_work_bg} provides the necessary background and related work. Sec.~\ref{sec:meas_setup_ch} describes all identified challenges in measurement setup for TD detection.  Sec.~\ref{sec:wehe} describes the Wehe tool and its mechanisms in the context of identified challenges, Sec.~\ref{sec:wehe_valid} provides the validation results. Our proposed method is described in Sec.~\ref{sec:wehe_improve} and Sec.~\ref{sec:conclusion} gives the conclusions.

\section{ Related work and background}
\label{sec:rel_work_bg}

As net-neutrality regulations were hotly debated in many countries, several authors have proposed methods to detect any net neutrality violations or deliberate traffic discriminations in the Internet.  Recent surveys \cite{nn_survey_2018, nn_survey_2020} cover all the tools and techniques used by them. The literature is also rich with various forms of discrimination and its detection, like discrimination of content providers \cite{nnvd_cpdiscri_pls}, end-users \cite{nnvd_userdiscri_plr}, specific services like BitTorrent \cite{bttest}. Our focus in this work is on detecting discrimination of streaming services (both audio and video) that constitute a major portion of internet traffic and consume significant bandwidth. We discuss some of the tools and techniques that is most relevant to our work.

\subsection{Existing TD detection tools}
There are two commonly used techniques for detecting TD in users' Internet traffic. One type of approach passively monitors traffic \cite{nano}. In such cases, the end-result or TD result is not immediately available to the user. Instead, the tool provides the aggregated result of traffic differentiation over the given ISP. Another type of detection technique uses specially crafted probing traffic - called Active probes. It analyses network response to probing traffic to detect any anomaly. \cite{nano,chkdiff, netpolice, shaperprobe, packsen} describes measurement setups based on such active probing. It uses traffic parameters such as packet loss, latency, packet sequence, or pattern to identify network operation characteristics or detect anomalies. Some tool uses multiple types of probing traffics called active differential probes. While one probing traffic type undergoes standard network middle-box processing, the other probing traffic type is supposed to evade any traffic differentiation by being not amenable to middle-box processing. Typically, in such differential probe methods, one traffic type mimics characters of traffic of services accessed from the original server (like Youtube, Netflix).
The other is a reference or control traffic. In this method, network responses to both traffic streams are compared to check if performance of any stream suffered against the other. \cite{glasnost, bonafide, wehe} are examples of such probing techniques. The differential probe method uses a common server to stores the content of services of interest, and this content is collected from the original servers. The common server thus replays the content and mimics the characteristics that original servers impart in the transfer of their content, like maintaining timing relationships between data packets. In the differential probe, traffic from the replay server must be treated similarly to that treated from the original servers. When it comes to HTTPS traffic,  the current tools do not ensure that traffic from replay servers is treated as that originated from an actual server and our work fills this gap.

\begin{figure}[!t]
	\centerline{\includegraphics[scale=0.13]{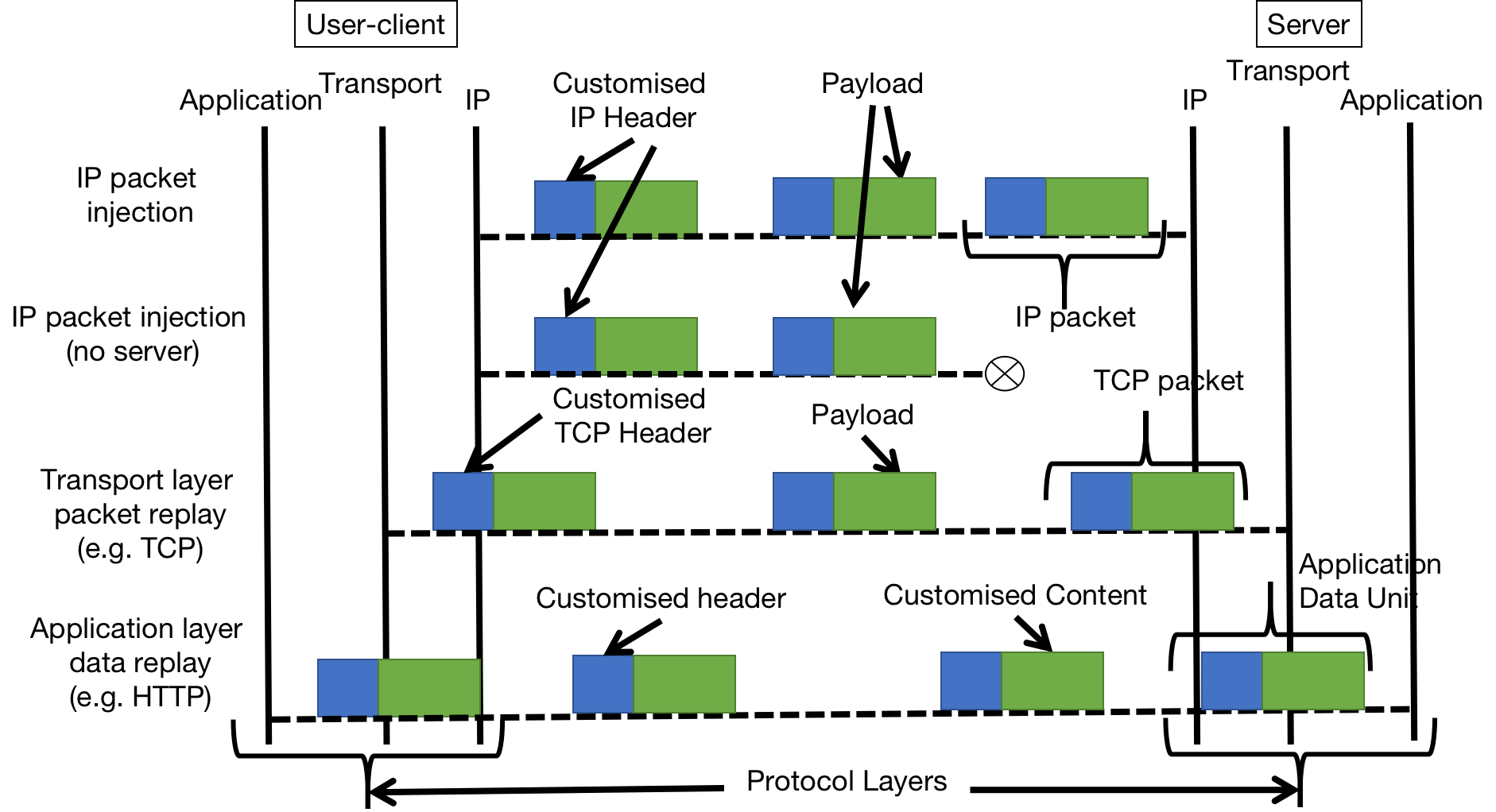}}
	\caption{Data replay techniques}
	\label{fig:data_replay}
\end{figure}

\begin{figure*}[!ht] 
	\centering
	\subfigure[Using internet browser\label{fig:gmail_intbrow}]{
		\includegraphics[width=0.49\linewidth]{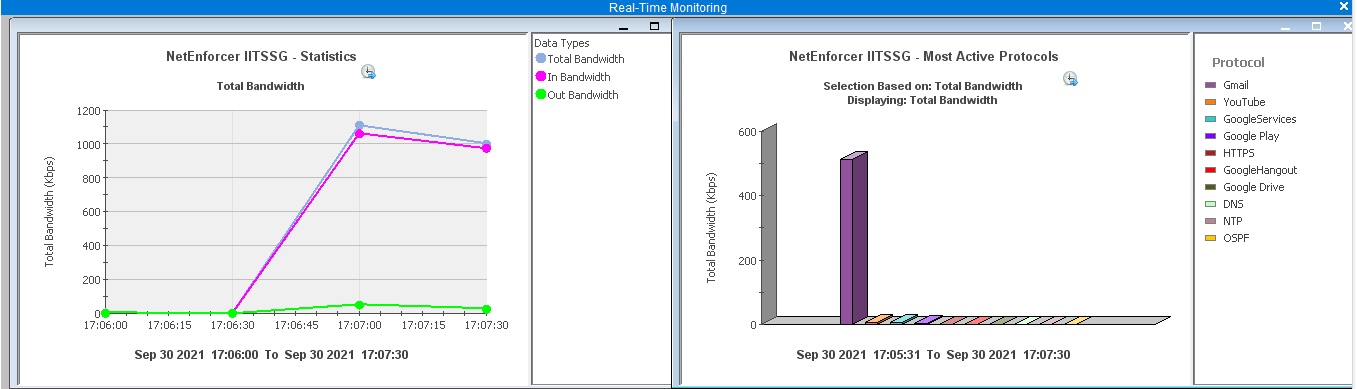}}
	\subfigure[Using user-client\label{fig:gmail_withoutsni}]{
		\includegraphics[width=0.49\linewidth]{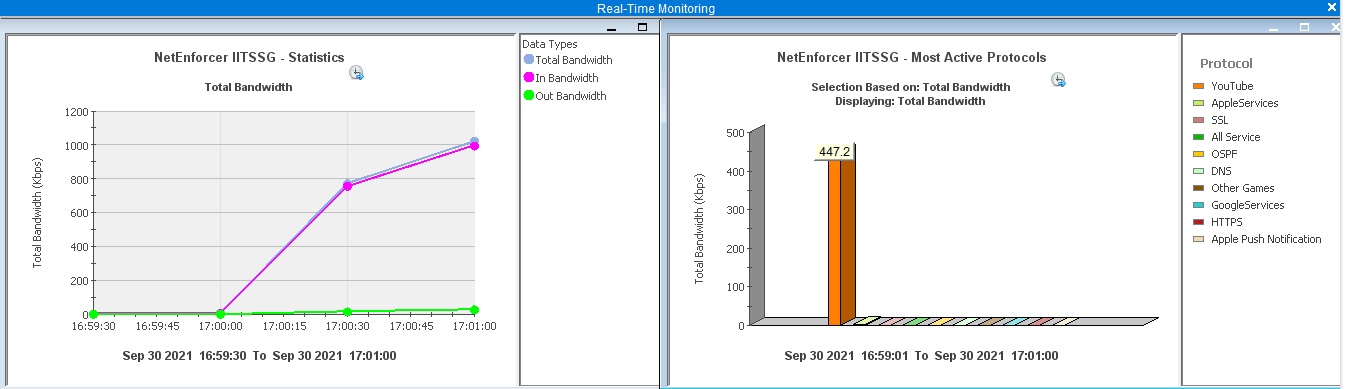}}
	\caption{Traffic classification of Gmail video play traffic with direct browser accesses and from user-client (TLS handshake not same as browser based access)}
	\label{fig:gmail_trcl}
\end{figure*}

\subsection{Traffic replay mechanisms}
\label{sec:trreplaymech}
The traffic replay mechanism mimics the client and server-side behavior for given application data exchange and the underlying protocol. There are many traffic replay tools available. \textit{Tcpreplay} \cite{tcp_replay} is one such replay tool that mimics the transport layer behavior for the given stream of transport layer packets.  Another example of a layer-specific replay is \textit{FlowrReplay} that runs at the application layer. The layer-specific replay tools are many times protocol dependent. The technique \textit{roleplayer} proposed in \cite{roleplayer} is capable of replaying application layer data in a protocol-independent manner. The replay layer selection (refer Fig.~\ref{fig:data_replay}) for traffic replay is crucial as it affects the receiver side's data collection as well as expected network response. The TCP layer replay adversely affects the traffic analysis as it requires special permission to collect traffic data for analysis.

 
\section{Challenges in TD detection measurement setup development}
\label{sec:meas_setup_ch}
A setup for detecting TD primarily consist of probing traffic generator, traffic data capturing system, and TD detection engine. In  this section we describe  challenges in engineering each of the components. 



\subsection{Network responses}
\label{sec:netresp}
The network response to the probing traffic is a fundamental input to the TD detection mechanism. The type of network response is dependent on the underlying methodology of the tool. Once fixed, the expected response from the network changes with the network configurations. Often, network nodes do not respond as expected to network management messages or do not classify the probing traffic in a specific manner. It happens either due to provisions in the associated Internet standard to deviate from the typical response or due to network policies for network resource management that are proprietary on which Internet standards do not have any control. 

Efficient network resource management is crucial for networks. It is exercised through various network middle-boxes across networks using network management practices (TMPs). These devices use either Deep Packet Inspection (DPI) \cite{dpi_survey} based or Deep Flow Inspection (DFI) \cite{dfi} based traffic classification for applying TMPs.

\begin{figure}[!ht]
    \centering
    \centerline{\includegraphics[scale=.09]{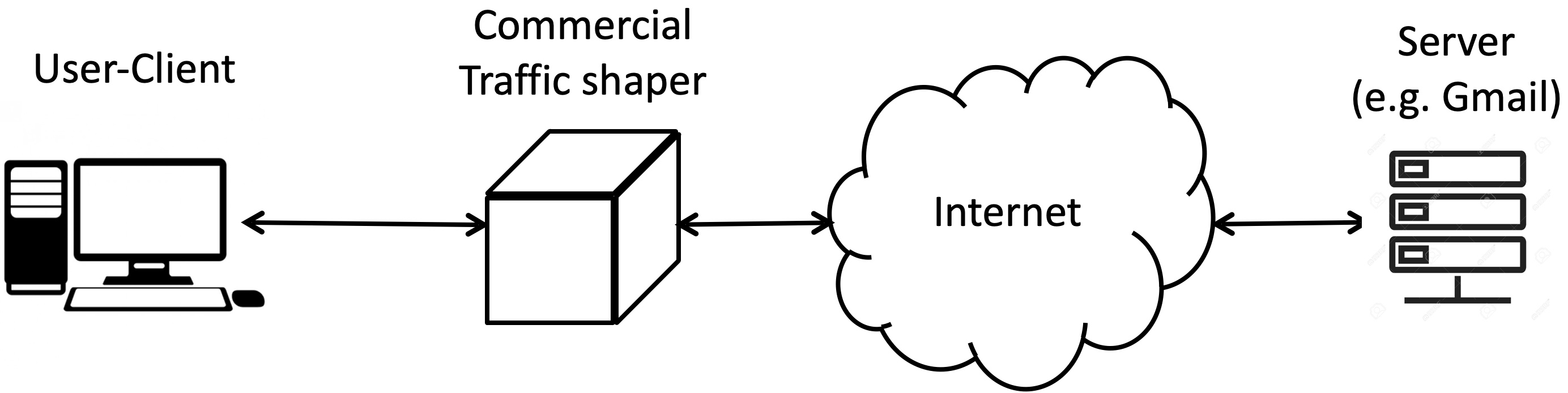}}
    \caption{Traffic classification validation}
    \label{fig:orgtrclclassvalid}
\end{figure}

We validated the traffic classification of traffic flows from the YouTube server using commercial traffic shaper as shown in Fig.~\ref{fig:orgtrclclassvalid}. Our validation setup consists of a user-client capable of downloading the content from a given service's original server (e.g. Gmail server). The traffic shaper is placed in-line between the user-client and YouTube server for validating the traffic classification of data traffic flow. 
The traffic classification of traffic flow using user-client and accessing it using internet browser are recorded using commercial traffic shaper's user interface.

Fig.~\ref{fig:gmail_trcl} shows the outcome of experiments as visible in commercial traffic shaper's user interface. The experiments are performed using an internet browser and user-client without performing TLS handshake same as that of performed by Internet browser. The traffic shaper user-interface window shows the internet traffic activity for last $10$ mins in the left window and service-wise classification of internet traffic in the right window. The commercial traffic shaper successfully detects the Gmail traffic for experiments using the internet browser (Fig.~\ref{fig:gmail_intbrow}). However, it could not detect any Gmail traffic for the experiment using user-client. The traffic shaper rather wrongly detects Gmail's traffic as YouTube traffic in the latter experiment ( Fig.~\ref{fig:gmail_withoutsni}). 

It indicates that commercial traffic shaper is always able to classify the Internet traffic correctly if accessed using Internet browser. However, the traffic classification is inconsistent or sometimes wrong in traffic characteristics based classification. As mentioned earlier, the traffic classification governs the network response. Thus generating a network response similar to that of the original service is a challenging task. . 

\subsection{TD Detection}
\label{sec:tdetection}
The TD detection algorithm is the core engine of the measurement setup for TD detection. Most of the time, it needs a specific type of input for its proper operations that is derived from the observed network response. The average throughput curves of probing traffic or specific traffic characteristics of sequence of network management response packets such as inter-packet times are examples of input information. TD detection involves the comparison of such specific network response across different traffic streams. However, these network response may vary across services due to different reasons without being subjected to deliberate traffic differentiation. 

The end-to-end connection between client and server for the Internet services is not dedicated. The best-effort nature of the IP layer packet forwarding results in packets from the same traffic stream to take different paths having different congestion environment. The performance comparison of streams experiencing different congestion is not reliable. 

Internet services employ various mechanisms to cope with the fluctuation in available bandwidth to provide a seamless end-user experience. Dynamic adaptive streaming over HTTP (DASH) is one such technique that modifies traffic characteristics such as speed or content characteristics such as coding rate. Each streaming service uses tailored techniques as per their requirements, and they are proprietary \cite{mdash}. Measurement setups such as passive monitoring systems face this challenge of normalizing various streaming services' performances for their difference in bandwidth fluctuation coping techniques. While active probing method overcome this issue using customised DASH or traffic replay, their probing traffic tends to saturates the available bandwidth, similar to point-to-point (p2p) traffic. Such traffic streams may lose their relevance as original service traffic.

The network middle-boxes applies QoS-based traffic management practices (TMPs) to efficiently allocate network resources across different types of services while maintaining the their bandwidth requirement. Thus different types of services may be allocated different bandwidths depending on the current network load and one has to take into account various confounding factors. 

\begin{figure}[htbp]
	\centering
	\includegraphics[scale=0.2]{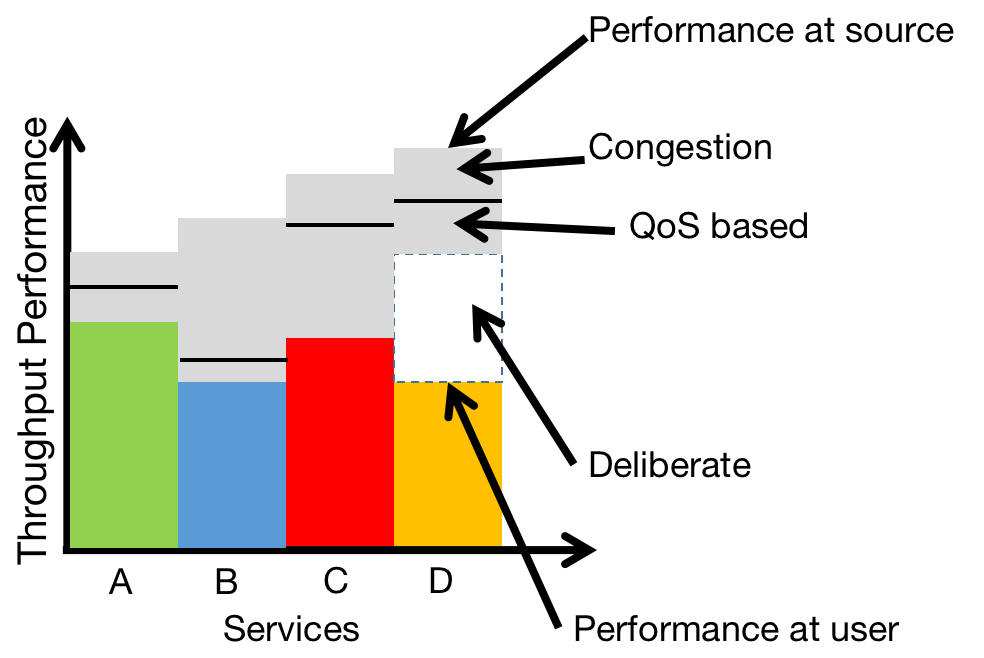}
	\caption{Variation in performance of services while traversing the Internet}
	\label{fig:perf_services}
\end{figure}

\begin{figure}[htbp]
	\centerline{\includegraphics[scale=0.45]{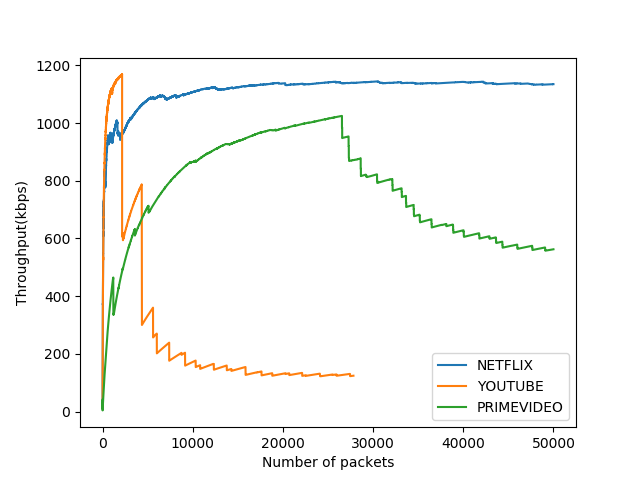}}
	\caption{Running avarage throughput for streaming services}
	\label{fig:th_str_services}
\end{figure}

Fig.~\ref{fig:perf_services} shows the effect of various confounding factors on the performances of Internet services (named A,B,C,D). In  the figure,  total height of each bar (consisting of the colored and grayed portion) shows throughput for each streaming service as governed by the transmission strategy (DASH in case of streaming services) of their servers.
The colored portion (other than grey) shows the throughput experienced at the client-side. The grey portion is the throughput lost in the network -- the upper part of the grayed portion shows the throughput lost due to network congestion and the lower grayed portion shows that lost due to network traffic management. Note that the server of each stream could be sending the traffic at different rates with their own proprietary DASH and that received by user-clients depend on the different levels of degradation the streams experience in the networks. The white portion in the throughput performance bar of service D shows throughput lost due to deliberate degradation. Identifying this discrimination of stream D just by comparing the throughout experienced at the user-client will be misleading due to non-uniformity in the transmission rates at the source and network effects.
We validated this aspect using three different streaming services namely Netflix, YouTube, and PrimeVideo. We plotted the average running throughput for these services for their traffic as shown in Fig.~\ref{fig:th_str_services}. It can be seen that each service is fetching the data differently depending on their congestion environment and server's data transmission schemes i.e. DASH. Thus the performances of these services are comparable without normalising the effect of confounding factors. 

The normalisation of effects of all confounding factors poses challenge in case of TD detection based on comparing network responses across services.

\subsection{Other challenges}
Internet services use a specific port number for communication. It is as per port reservations defined in Internet standards \cite{serv_port}, e.g., port $80$ for HTTP traffic and $22$ for SSH (Secure Shell ) traffic. Thus the port number used in the transmission of probing data plays a vital role in traffic classification by network middle-boxes \cite{cisco-sdavc}. Using correct data to be used on the pre-assigned port number for a given service and making it as authentic as original service's traffic stream is a challenging task. It requires a thorough understanding of network traffic classification on that port.

\section{Case Study : Wehe - TD detection tool for mobile environment}
\label{sec:wehe}
Wehe \cite{wehe} is a tool for the detection of traffic differentiation over mobile networks (e.g., cellular and WiFi). It is available as an App on Android and the iOS platform. The user database of the Wehe tool consists of 126,249 users across 2,735 ISPs in 183 countries/regions generating 1,045,413 crowd-sourced measurements. European national telecom regulator, the US FTC and FCC, US Senators, and numerous US state legislators have used the Wehe tool's findings. 

The tool supports TD detection for many popular services such as Netflix, YouTube. The tool runs TD detection tests by coordinating with its server, called the ``replay server". The replay server keeps track of active users and maps replay runs to the correct user's service.

\subsection{Traffic generation}
Wehe uses the ``record-and-replay" method for generating probing traffic. The user-client exchanges the probing traffic with the replay server as per the replay script that captures the application-level traffic behavior, including the port number, data sequence, and timing dependencies from the original service's network logs. Preserving timing is a crucial feature of Wehe's approach. It expects network devices to use this information in case of non-availability of any other means to classify applications, e.g., HTTPS encrypted data transfer. 

Wehe tool uses two types of probing traffic streams. While one stream (\textit{original} replay) is the same as the original service's application-level network trace, another traffic stream (\textit{control} replay) differs substantially from the first traffic stream. In one approach, Wehe uses the VPN channel to send a second probing traffic stream. This approach uses the meddle VPN \cite{meddle} framework for data transfer and server-side packet capture. Another approach uses the bit-reversed version of the first traffic stream sent one the same channel. Currently, Wehe uses the latter approach due to its superior results. 

\subsection{Over the network response}
Wehe is a differential detector tool that compares the network responses for two types of traffic streams generated by the tool. The service-specific information present in the \textit{original} replay is useful for network devices with Deep Packet Inspection (DPI) capability to identify and classify the service correctly. So, the \textit{original} replay's traffic performance over the Internet closely resembles the original application traffic on the same network. While \textit{original} replay is exposed for detection to network devices, the traffic streams with bit reversed data or \textit{control} replay is equally ``not detectable" for classification. Thus it is expected that the \textit{control} replay traffic evades the content-based application-specific traffic differentiation. The performances of two such traffic streams (detectable and non-detectable) differ if network devices apply different traffic management or traffic differentiation on each traffic stream as per content-based classification. 
   
\subsection{TD detection scenario expectations}
Wehe compares the throughput performances of \textit{original} replay and \textit{control} replay to detect TD. The methodology uses the throughput as a comparison metric due to its sensitivity to bandwidth-limiting traffic shaping. However, the tool expects that the TD detection algorithm does not detect TD based on throughput for traffic streams with traffic rates below the shaping rate. The rationale is that the shaper cannot affect the performance of such an application stream. 

Many times both traffic streams get affected by other factors such as signal strength, congestion. It creates an irregularity in the received performance due to bandwidth volatility. It is mentioned to be leading to incorrect differentiation detection. The tool performs multiple test replays to overcome the effect of bandwidth volatility. 

\subsection{Operational requirements}
Wehe server needs side-channels for each client to associate it with precisely one app replay. This side-channel supplies information about replay runs to the server. Each user directly connected to the Wehe replay server is uniquely identifiable on the server-side with an associated IP address of side channels that maps each replay to exactly one App. 

The other operational requirement is that the Wehe client-server communication uses customized socket connections with specific keep-alive behavior. Sometimes, the usage of translucent proxies by the user-client modifies this behavior. The Replay server handles this situation by handling such unexpected connections. The protocol-specific proxies, e.g., HTTP proxy, connect the user-client to the server through itself using specific port numbers, e.g., 80/443 for HTTP/HTTPS. Nevertheless, it allows the user-client to connect to the server for connections using other protocols directly. The side channels of Wehe do not use HTTP/HTTPS connections. So the IP address for the same user differs for side-channel and replay runs. Wehe server detects such connections and indicates such connections to the user-client using a special message. The special message triggers the exchange of further communication with a customized header. 

\begin{figure*}[!ht] 
	\centering
	\subfigure[Using internet browser\label{fig:yt_replay_intbrow}]{
		\includegraphics[width=0.49\linewidth]{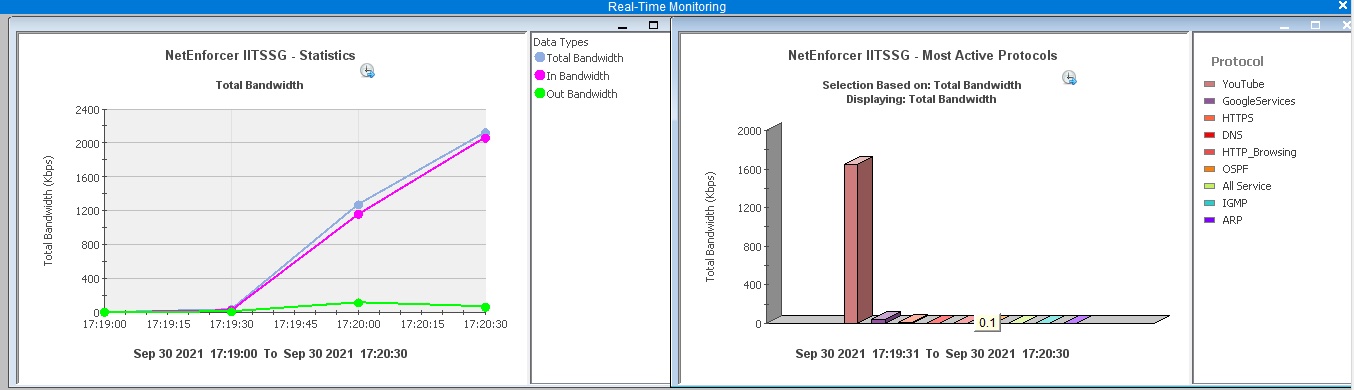}}
	\subfigure[Using user-client \label{fig:yt_replay_withoutsni}]{
		\includegraphics[width=0.49\linewidth]{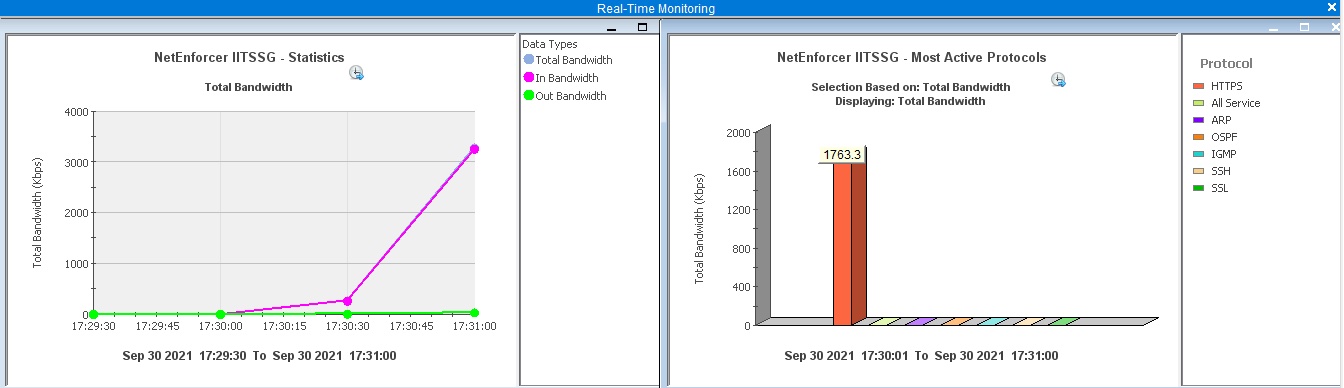}}
	\caption{Traffic classification of replayed YouTube traffic}
	\label{fig:replay_trcla}
\end{figure*}

\subsection{Challenges of validating Wehe}
 The TD mechanism of Wehe is straightforward to use — the requirement changes when using it for its validation. The validation process may need to launch only one type of replay for different services during one test or may need to launch all replays in parallel. These are not requirements related to TD detection, Wehe's primary goal, so understandably not supported. Hence the validation of Wehe's working in such scenarios needs a specific client-server setup. Here the challenge is to separate the intended scenario-specific Wehe's mechanism so that the resulting system still mimics Wehe's actual behavior. 

Moreover, Wehe does not provide error/failure notifications in all scenarios. Instead, it prompts the user to reopen the App. As a result, the validation setup loses vital feedback information regarding the error/failure induced by its validation scenario.

\section{Shortcoming  of Wehe on HTTPS traffic}
\label{sec:wehe_valid}
Our study focuses on validating the network responses for the replayed traffic streams, TD detection, and operational feasibility in various network configurations. While operational feasibility is validated using the publicly available ``Wehe" Android app on Google Playstore, TD detection scenarios are validated using theoretical arguments. The validation of network responses requires bandwidth analysis of the received traffic stream. This analysis requires the network logs for the specific replay performed as per the validation scenario. The replay done on the device and multiple other streaming services running in parallel is one such scenario. Wehe app does not immediately provide such network logs for the replays after the completion of tests. So, we implemented the user-client and server that mimics the behavior of the Wehe tool. 


We use a client-server setup similar to the setup shown in Fig.~\ref{fig:orgtrclclassvalid} for validating Wehe. In the current setup, we replaced the original service's server with replay server. The user-client and replay server are connected through a commercial traffic shaper.  Moreover, our setup has a provision to perform multiple replays in parallel. Our validation setup does not need administrative channels and overheads, e.g., side-channels. Our server always needs to support a single-user client. The validation of scenarios with multiple clients uses the Wehe App directly due to the non-requirement of associated traffic analysis.

Reminder of this section describes results of the validation.

\begin{figure*}[!ht] 
	\centering
	\subfigure[Only Wehe\label{fig:wehe_th}]{
		\includegraphics[width=0.3\linewidth]{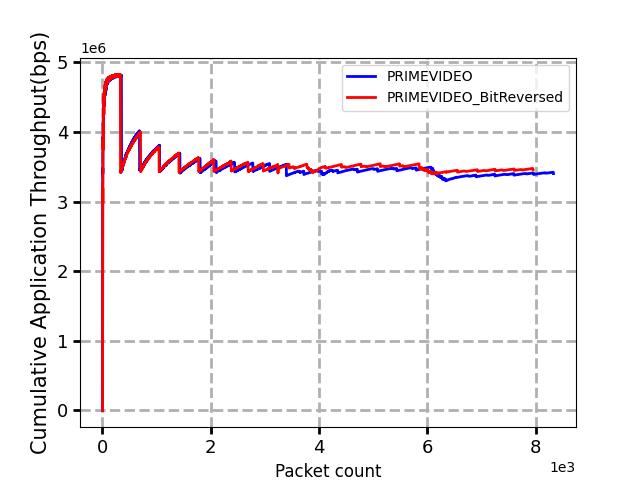}}
	\subfigure[Wehe plus one service\label{fig:wehe_p1}]{
		\includegraphics[width=0.3\linewidth]{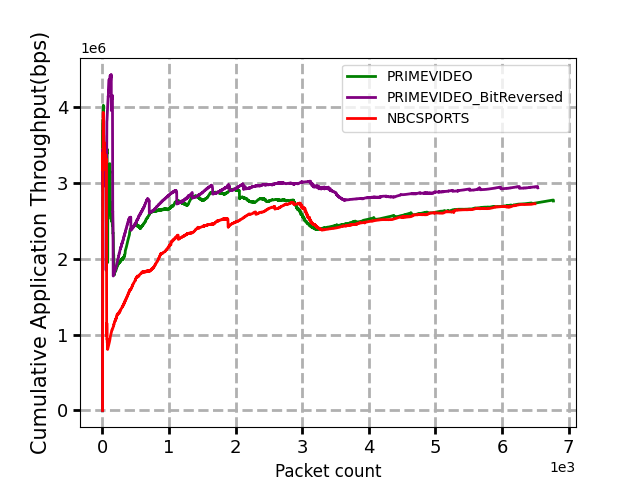}}
	\subfigure[Wehe plus two services\label{fig:wehe_p2}]{
		\includegraphics[width=0.3\linewidth]{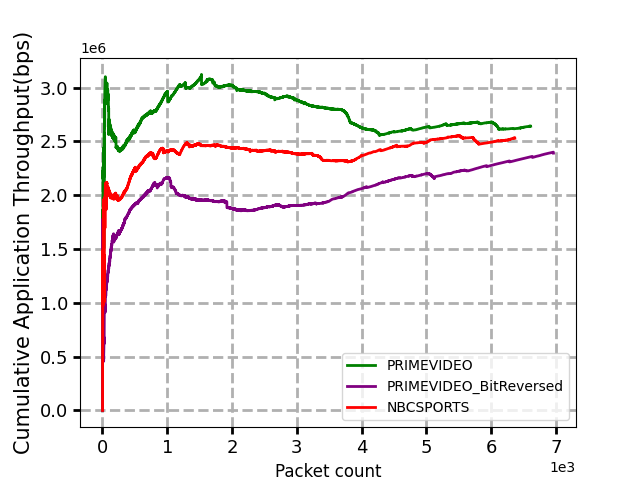}}
	\caption{Effect of network load on Wehe's traffic stream performances}
	\label{fig:wehe_th_ps}
\end{figure*}

\subsection{Original service's traffic emulation}
\label{sec:wehe_org_serv_tr_emul}
The network responses are dependent on the application of network policies based on correct probing traffic classification by network middle-boxes as mentioned in Sec.\ref{sec:netresp}. We validated the classification of Wehe's emulated traffic using a commercial traffic shaper. The classification of emulated traffic is observed using the user interface of commercial traffic shaper. 

For performing experiment, the YouTube application data is replayed from replay server to user-client through commercial traffic shaper. During data transfer, the commercial traffic shaper's user-interface window is observed for the presence of YouTube traffic. We also accessed the YouTube traffic using an internet browser and observed the traffic shaper's classification outcome to baseline our traffic classification observations.

Fig.~\ref{fig:replay_trcla} shows the traffic classification outcome as shown by commercial traffic shaper's user-interface window for traffic accessed directly using an internet browser from a YouTube server. It shows the internet activity in the left window and the classification of corresponding traffic in the right window. 

Fig.~\ref{fig:yt_replay_intbrow} shows that traffic accessed using internet browser is correctly classified as YouTube. This is inline with commercial traffic shaper's intended behaviour.

Fig.~\ref{fig:yt_replay_withoutsni} shows the traffic classification outcome for traffic accessed using user-client. It shows the evidence of no YouTube traffic transferred over the communication link. Moreover, it classifies the same traffic as HTTPS traffic. The outcome of this experiment shows that not all network middle-boxes can correctly classify the Wehe's replayed traffic.

\subsection{Effect of data rate in replay script}
\label{subsec:td_org_replay} The probing traffic generation impacts the network response as expected by the TD detection algorithm
Wehe uses the traffic trace from the original service for generating replay scripts that preserve the application data and its timing relationship. This replay script is used over the original network and also on networks that are differently geo-located. As traffic shaping rate varies across networks for the same service (as mentioned in \cite{wehe_res_2019}), the traffic rate preserved in the replay script can be different from the traffic shaping rate of the currently considered network. The replay traffic rate can be lower than the traffic shaping rate.

The Wehe methodology does not detect traffic differentiation if the replay script's traffic rate is lower than the network's shaping rate as it does not affect the traffic stream. Such replay scripts can never detect traffic shaping on such networks. Thus Wehe tool's TD detection capability is limited by the replay script's probing traffic rate.

\subsection{Usage of port number 80}
\label{subsec:port_80}
The network responses are driven by network middle-boxes perception about the probing traffic (refer Sec.~\ref{sec:netresp}). The replay script preserves the data in the applications' original network trace.  The original application servers use port $80$ for the plain-text data and port  $443$ for encrypted data transfer. Wehe replay script directly uses the encrypted data from the application's network trace and transmits it on port $80$. In such cases, Wehe expects its \textit{original} replay traffic stream to be classified correctly by network devices using encrypted application data. It is impossible for such data on port  $80$ as encrypted traffic data cannot expose its identification to the network device. Thus Wehe tool cannot generate the required traffic streams for services running on the port number $443$ due to default usage of port $80$ for replay run.

\subsection{Traffic load governed network behavior}\label{subsec:tr_load}
Note that scarcity of resources prompt networks to apply certain network traffic management, especially in heavy network load scenarios, that are beneficial for all active services on its network, e.g., QoS-based traffic management (refer Sec.~\ref{sec:netresp}). We validated the effect of such traffic management on the performances of both \textit{control} and \textit{original} replays. The validation uses the following three scenarios for the validation,
\begin{itemize}
	\item Replaying only Wehe's two traffic streams without any load on the network (Fig~\ref{fig:wehe_th})
	\item Replaying Wehe's three traffic streams with one additional streaming service running in parallel (Fig.~\ref{fig:wehe_p1})
	\item Replaying Wehe's three traffic streams with $2$ additional streaming services running in parallel (Fig.~\ref{fig:wehe_p2})
\end{itemize}
The performances in Fig.~\ref{fig:wehe_th} show that performances of traffic streams generated by the Wehe tool are the same under no additional network load conditions. As network load increases, the performance of \textit{control} replay deviates from that of \textit{original} replay and at higher level (Fig.~\ref{fig:wehe_p1}). While performance of \textit{control} replay further deviates from \textit{original} replay on lower side, two \textit{original} replays still shows similar performances as shown in Fig.~\ref{fig:wehe_p2}. It invalidates the Wehe tool's expectation of \textit{control} replay not getting differentiated. It also invalidates the claim of the tool of detecting the TD due to total bandwidth.  

\begin{figure*}[!ht] 
	\centering
	\subfigure[Gmail video play\label{fig:gmail_withsni}]{
		\includegraphics[width=0.49\linewidth]{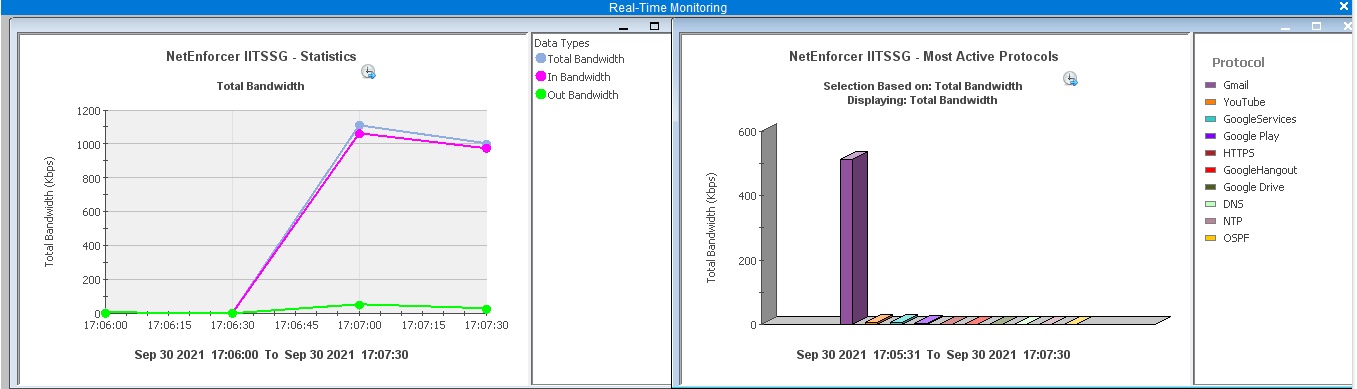}}
	\subfigure[YouTube replay traffic\label{fig:yt_replay_withsni}]{
		\includegraphics[width=0.49\linewidth]{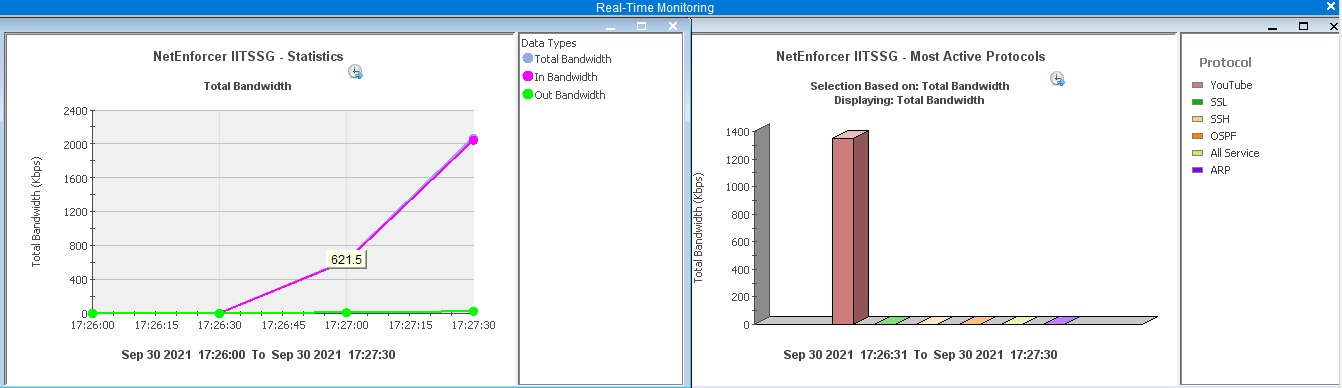}}
	\caption{Traffic classification of Gmail video play and YouTube replay traffic with appropriate SNI}
	\label{fig:trcl_withsni}
\end{figure*}

\subsection{Wehe tests from multiple devices within the same sub-net} 
\begin{figure}[htbp]
	\centerline{\includegraphics[scale=0.1]{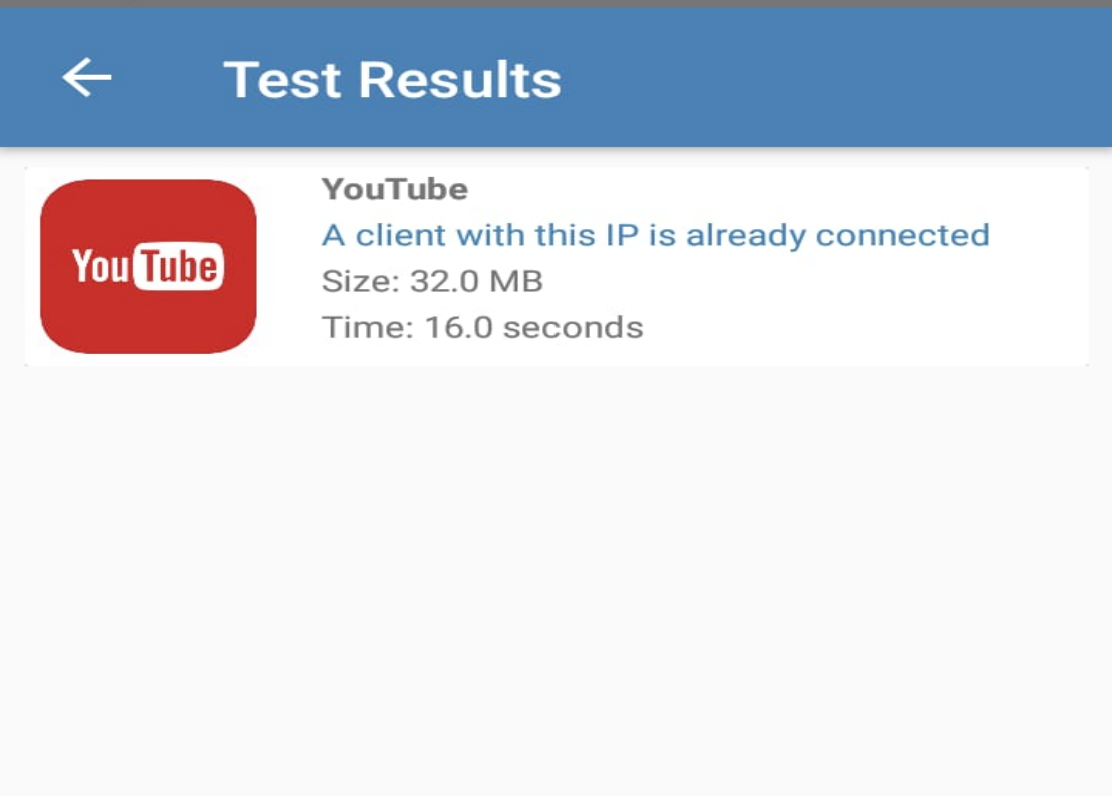}}
	\caption{Wehe apps running on multiple devices within the same subnet and testing different service}
	\label{fig:wehe_mul_device_error}
\end{figure}

The side-channels are introduced in Wehe design to support multiple user clients simultaneously. Side channels also assist in identifying the mapping between user-client and a combination of  IP addresses and ports.  It is useful in the case of networks using NATs \cite{nat_rfc3022}. We validated Wehe's support for multiple clients and NAT-enabled networks using two different tests.
First, we connected two user-clients from within the same subnet, i.e., clients sharing the same public IP address. In one test, the Wehe tool tests the same service on both devices, e.g., Wehe App on both devices tests for YouTube. The result shows that the Wehe test completed finishing on only one device while Wehe App abruptly closed on another device. We repeated the same scenario, but this time Wehe tests different services, e.g., Wehe on one device testing YouTube during another testing Netflix. We found that the Wehe test on one device completes properly while the Wehe test on another device throws an error on the screen, informing the user that another client is already performing the test, as shown in Fig.~\ref{fig:wehe_mul_device_error}. These tests show that Wehe does not support multiple devices if they share the same IP address.
While a side-channel is useful to identify each replay from a user-client connected directly to the Wehe replay server, it is not useful in the network using NAT devices. Multiple users share the same IP address in the case of NAT. In such cases, the side channel can not uniquely map each replay run to a client. It limits the usage of Wehe to only one active client per replay server and ISP and application. This limitation is documented by Wehe developers as well.

\subsection{Effect of device network load on TD detection}
\label{subsec:non_exhaust_bw} Wehe's replay server uses the same timings between application data transfer as that of original application traffic. Such a transmission strategy is expected not to exhaust available bandwidth. Hence the effect of source rate modulation due to overshooting of traffic rate above available bandwidth is unlikely. It makes, \textit{original} and \textit{control} replays show similar traffic performances unless deliberately modified by network policies i.e. TD. 

Nevertheless, this expectation is not always satisfied as the traffic data rate is modulated by the network load at the user device while performing Wehe tests. Such perturbations create discrepancy as the effect of time-varying current network load on the probing traffic is also time-varying and may not always be the same. The back-to-back replay strategy of Wehe ensures that both (\textit{original} and \textit{control} replay) probing traffic streams gets affected differently by the current network load. Under such network load on the device side, the notion of services not exhausting available bandwidth ceases to exist along with its benefits for TD detection. It is necessary to normalise such confounding factors (refer Sec.~\ref{sec:tdetection}) before TD detection.

\section{TD detection of HTTPS traffic}
\label{sec:wehe_improve}
As described in Sec.~\ref{sec:wehe_valid}, Wehe tool suffers from critical weakness of not achieving intended traffic classification of its \textit{Original replay} traffic due to encrypted nature of most of the modern streaming services as demonstrated in Sec.~\ref{sec:wehe_org_serv_tr_emul}. In this section, we propose a method to overcome this shortcoming  and make it useful to detect discrimination of HTTPS traffic.

The payload-based traffic classification techniques do not apply to encrypted traffic. Commercial traffic shapers primarily use the Server Name Indication or SNI-based traffic classification technique to overcome the classification of encrypted traffic. It is used in non-collaborative manner \cite{cisco-sdavc} with other classification techniques due to its higher accuracy \cite{trcl_survey}.

\begin{figure}[htbp]
	\centerline{\includegraphics[scale=0.2]{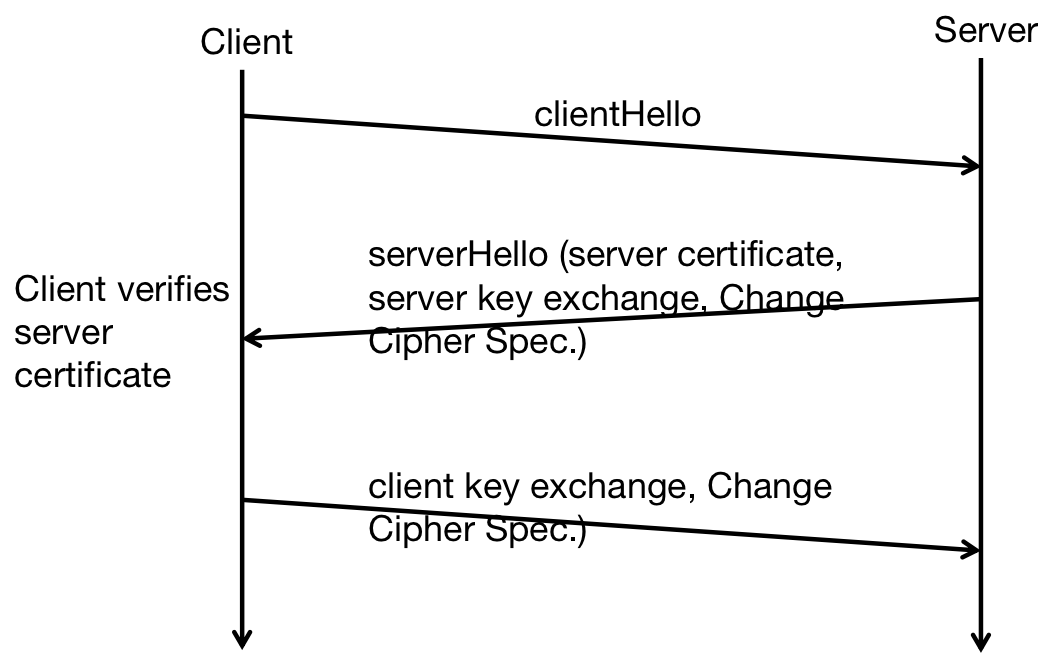}}
	\caption{TLS handshake sequence}
	\label{fig:tls_hs}
\end{figure}

Transport layer security (TLS) \cite{tls_rfc7858} is an Internet protocol that provides channel security to transport layer protocol communication. The typical TLS handshake is as shown in Fig.~\ref{fig:tls_hs}. SNI is a parameter in the initial TLS handshake message named ``ClientHello". It is used on the server-side to select service-specific SSL certificates. Thus, it uniquely identifies the service. ``ClientHello" is the first handshake message, its contents are unencrypted including SNI. Almost all streaming services use TLS for secure channel establishment. Thus making SNI-based traffic classification applicable to almost all streaming services. So SNI is the primary parameter used by traffic shapers to classify traffic exchanged over the public internet using standard HTTPS port $443$ i.e., encrypted traffic.  

Before demonstrating the usability of SNI in traffic classification of encrypted traffic for Wehe replay traffic, we first validated the importance of SNI parameters in the classification of traffic from YouTube's original server. We used the setup as shown in Fig~\ref{fig:orgtrclclassvalid}. We repeated the experiment in Sec.~\ref{sec:netresp} with user-client performing TLS handshake using appropriate value of SNI parameter extracted from original service's network logs. The Fig.~\ref{fig:gmail_withsni} shows the outcome of the experiment as the correct classification of Gmail traffic. Note that the same traffic was wrongly classified as YouTube traffic when transferred without SNI (refer Fig.~\ref{fig:gmail_withoutsni}).

Based on the above validation, we propose to use SNI in Wehe's replay traffic to overcome its shortcoming of correct replay traffic classification. To validate this point, we repeated the experiment in Sec.~\ref{sec:wehe_org_serv_tr_emul} with appropriate SNI (SNI extracted from original YouTube service's traffic) used in the TLS handshake. We used the experiment setup as used in Sec.\ref{sec:wehe_valid}. The Fig.~\ref{fig:yt_replay_withsni} shows the outcome of this experiment. As seen, that the traffic shaper has correctly detected YouTube replay traffic. This observation establishes that using service-specific SNI in TLS handshake messages leads to the correct classification of replay traffic. 

The Wehe tool employing our new SNI-based mechanism to emulate the original service's traffic can detect discrimination of HTTPS traffic and also work with a larger set of network middle-boxes, including those that fail to classify replay traffic correctly based on traffic characteristics alone.

\section{Conclusion}
\label{sec:conclusion}
Net neutrality violation detection is a need of an hour. As many of the ISPs are also content providers these days, they compete with each other, which can lead to one deliberately discriminating the services of the other to gain market share. However, users should have the freedom to choose services as per their wishes. Our work considered various challenges in the detection of traffic discrimination in HTTPS traffic. 

As a case study, we validated Wehe, one of the latest tools available to detect traffic differentiation. The described challenges helped us divide the entire tool into multiple interdependent components and validate them independently. Our validation using commercial traffic shaper revealed that traffic in Wehe setup may not mimic the characteristics of HTTPS traffic accessed from the original servers. Hence, middle-boxes may not subject them to intended discrimination. Thus, Wehe may not detect discrimination of HTTPS traffic. Our new method that uses the appropriate SNI parameter value in the initial TLS handshake message overcomes this shortcoming. Hence our work provided a mechanism to detect a wide range of possible discriminations on the Internet.

\newpage

\bibliographystyle{IEEEtran}
\bibliography{IEEEabrv,vinod_bibfile}

\end{document}